\documentclass[aps,reprint,prl,groupedaddress]{revtex4-2}
\usepackage{xcolor}
\usepackage{float}
\usepackage{amsmath,amsfonts,amssymb,epsfig,graphicx}
\usepackage{slashed} 
\usepackage{hyperref}
\usepackage[normalem]{ulem}
\usepackage{lineno}
\graphicspath{{./pic/}}
\begin{document}

\title{Probing darK Matter Using free leptONs: PKMUON}

\author{Alim \surname{Ruzi}}
\email[]{alim.ruzi@pku.edu.cn}
\affiliation{State Key Laboratory of Nuclear Physics and Technology, School of Physics, Peking University, Beijing, 100871, China}

\author{Chen \surname{Zhou}}
\email[]{czhouphy@pku.edu.cn}
\affiliation{State Key Laboratory of Nuclear Physics and Technology, School of Physics, Peking University, Beijing, 100871, China}

\author{Xiaohu \surname{Sun}}
%\email[]{Xiaohu.Sun@pku.edu.cn}
\affiliation{State Key Laboratory of Nuclear Physics and Technology, School of Physics, Peking University, Beijing, 100871, China}

\author{Dayong \surname{Wang}}
%\email[]{dayong.wang@pku.edu.cn}
\affiliation{State Key Laboratory of Nuclear Physics and Technology, School of Physics, Peking University, Beijing, 100871, China}

\author{Siguang \surname{Wang}}
%\email[]{siguang@pku.edu.cn}
\affiliation{State Key Laboratory of Nuclear Physics and Technology, School of Physics, Peking University, Beijing, 100871, China}

\author{Yong \surname{Ban}}
\email[]{bany@pku.edu.cn}
\affiliation{State Key Laboratory of Nuclear Physics and Technology, School of Physics, Peking University, Beijing, 100871, China}

\author{Yajun \surname{Mao}}
\email[]{maoyj@pku.edu.cn}
\affiliation{State Key Laboratory of Nuclear Physics and Technology, School of Physics, Peking University, Beijing, 100871, China}

\author{Qite \surname{Li}}
\email[]{liqt@pku.edu.cn}
\affiliation{State Key Laboratory of Nuclear Physics and Technology, School of Physics, Peking University, Beijing, 100871, China}

\author{Qiang \surname{Li}}
\email[]{qliphy0@pku.edu.cn}
\affiliation{State Key Laboratory of Nuclear Physics and Technology, School of Physics, Peking University, Beijing, 100871, China}

\begin{abstract}
We propose a new method to detect sub-GeV dark matter, through their scatterings from free leptons and the resulting kinematic shifts. Specially, such an experiment can detect dark matter interacting solely with muons. The experiment proposed here is to directly probe muon-philic dark matter, in a model-independent way. Its complementarity with the muon on target proposal, is similar to, e.g. XENON/PandaX and ATLAS/CMS on dark matter searches. Moreover, our proposal can work better for relatively heavy dark matter such as in the sub-GeV region. We start with a small device of a size around 0.1 to 1 meter, using atmospheric muons to set up a prototype. Within only one year of operation, the sensitivity on cross section of dark matter scattering with muons can already reach $\sigma_D\sim 10^{-19 (-20,\,-18)}\rm{cm}^{2}$ for a dark mater $\rm{M_D}=100\, (10,\,1000)$ MeV.
We can then interface the device with a high intensity muon beam of $10^{12}$/bunch. Within one year, the sensitivity can reach $\sigma_D\sim 10^{-27 (-28,\,-26)}\rm{cm}^{2}$ for $\rm{M_D}=100\, (10,\,1000)$ MeV. 
\end{abstract}

\maketitle

\section{Motivations} 
The Standard Model (SM) of particle physics does not explain some experimental observations, like the neutrino mass, and Dark Matter (DM). Physics beyond SM is required to include some hypothetical particles as well as new interactions. Among the hypothesized particles, the nature of DM is currently a major question for both cosmology and particle physics. Although we believe in its existence in the Universe based on some cosmological observations, no experiments have directly observed any kind of DM particles. There are several experiments dedicated to the direct or indirect detection of light DM particles, whose mass ranges from sub-MeV to GeV scale. The paradigm of WIMPs, Weakly Interacting Massive Particles, is well studied in the DM sector and many experiments were conducted to search for them, yet no observations of WIMPs have been made. However, it is possible that WIMP-related models have been misleading. So it is necessary to explore other theoretically motivated scenarios, for example, low mass DM, or muon-philic DM~\cite{Essig:2022dfa,Harris:2022vnx,Bai:2014osa}. 

Detection of sub-MeV DM is a more difficult task and current limits are much weaker than those for the GeV region. Traditional direct searches for DM, looking for nuclear recoils in deep underground detectors, are challenging due to insufficient recoil energy, and the experimental searches of cold sub-GeV DM have focused on the Migdal effect~\cite{migdal} and the interaction with electrons~\cite{EDELWEISS,SENSEI}. Low-mass DM searches with liquid Helium has also been proposed~\cite{Liao:2021npo,Liao:2022zqg}. Recently, experiments also exploit the fact that a fraction of the cold DM is boosted to relativistic energies and thus can be efficiently detected in direct detection experiments~\cite{Super-Kamiokande:2022ncz,PandaX-II:2021kai,Bringmann:2018cvk,Plestid:2020kdm,Hu:2016xas}. More projections on low-threshold DM direct detection in the next decade can be found in Ref.~\cite{Essig:2022dfa,Elor:2021swj}.

On the other hand, lepton beams may play an important role in the detection of light DM particles.
There are some proposals using electron beams for direct detection of sub-GeV DM particles~\cite{Essig:2012yx,Battaglieri:2020lds,Berlin:2020uwy,LDMX:2018cma,DarkSide:2022knj}. These approaches utilize electron beams to probe invisibly-decaying particles which couple to electrons.
Some theoretical models like 
 Muonphilic DM and Lepton portal DM~\cite{AlAli:2021let,Bai:2014osa} show that DM states only couple to the charged leptons, and predominantly interact with muons with a new type of interaction called muonic force. The quest for explaining the pronounced muon magnetic moment anomaly from the Fermilab Muon g-2 collaboration with 4.2 $\sigma$ discrepancy~\cite{Muong-2:2021ojo} drives people to search such DM preferentially interacting with muons through proton beam-dump experiments utilizing muon beams~\cite{Forbes:2022bvo}. The muonphilic DM scenario may reconcile this anomaly by introducing a force carrier particle, muonphilic bosons~\cite{Gninenko:2014pea}.
 
Accelerator muon beams have been proposed as a potential tool for detecting DM particles by NA64-$\mu$ experiment~\cite{Chen:2018vkr} as well as $M^3$~\cite{Kahn:2018cqs} and FNAL-$\mu$ experiment at Fermilab~\cite{Chen:2017awl}, and also for the future muon collider~\cite{Cesarotti:2022ttv}. The idea behind is to look for the muon deflection caused by scattering with DM as well as the the energy loss pattern of muons.  

Here we propose a novel method to directly detect light mass DM through its scattering with abundant atmospheric muons or accelerator beams, to directly probe muon-philic DM in a model-independent way. Its complementarity with the $M^3$ proposal~\cite{Kahn:2018cqs}, is similar as, e.g. XENON/PandaX and ATLAS/CMS on DM searches~\cite{Buchmueller:2017qhf}.  $M^3$ is a two-phase experiment, in model-dependant way, aiming for the detection of sub-GeV DM  through the sensitivity to muon g-2 in phase 1 and $U(1)_{L_{\mu}-L_{\tau}}$ thermal DM in phase 2, respectively. The core idea in $M^3$ proposal is to identify missing momentum of the muons after scattering with a fixed nuclear target. It is assuming that a scalar or vector type mediator is radiated off muons and eventually decays invisibly into DM or neutrinos.

The sensitivity reach of our proposal, which is comparable to that of the $M^3$ experiment, is given in detail in the following sections.

\section{Our proposal} 

\begin{figure}
    \centering
    \includegraphics[width=1.\columnwidth]{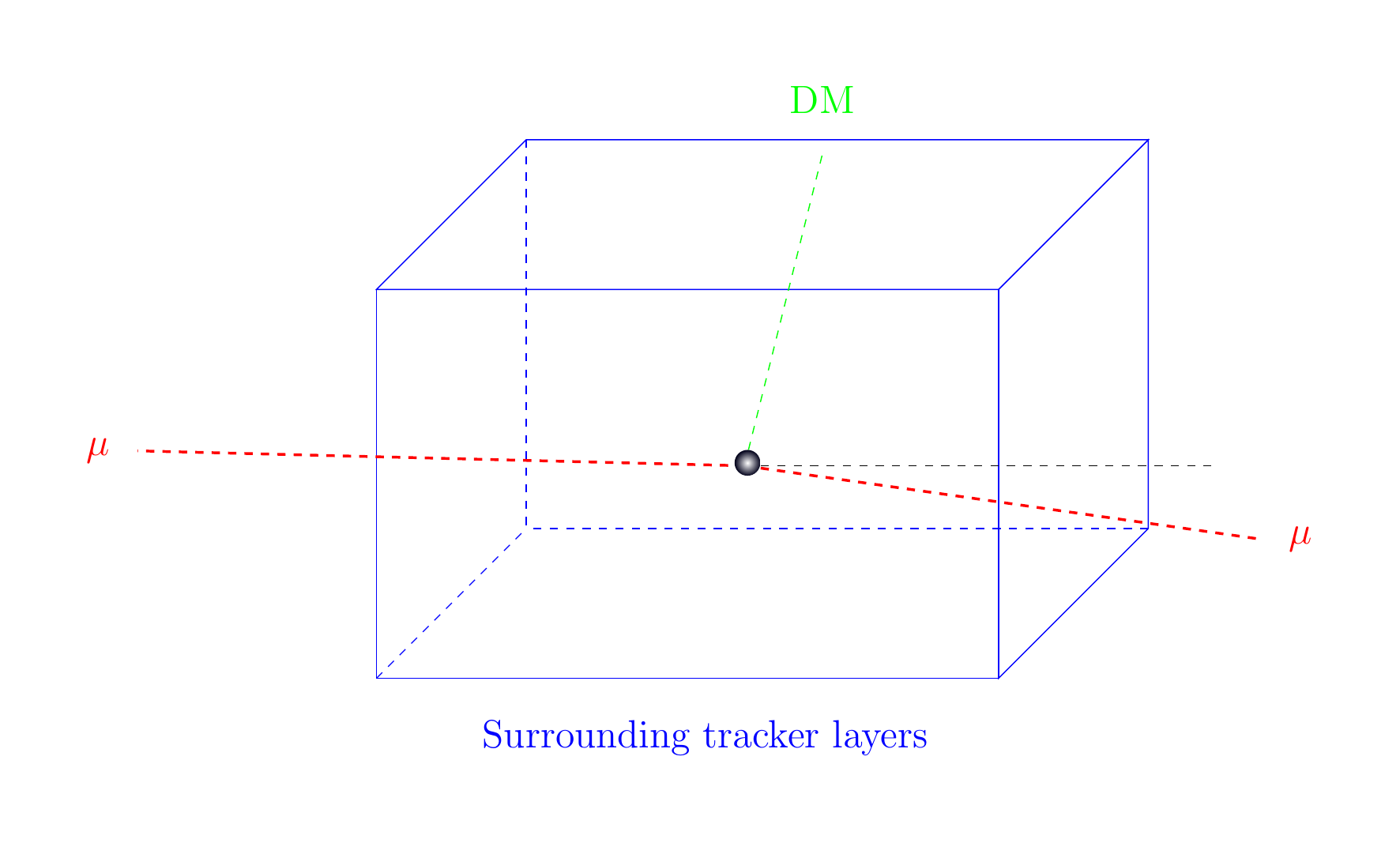}
    \caption{Illustration of experiments to detect low mass DM with free leptons. The resulting kinematic shifts of leptons kicked by DM can be measurable with tracking detectors surrounding a vacuum region. A veto region along the chamber can be defined based on the cross-point of the in and out tracks to suppress backgrounds. }
    \label{fig:eon}
\end{figure}

In this paper, we are interested in a novel method to detect low-mass lepton-philic DM with free leptons (electrons or muons), as shown in Fig.~\ref{fig:eon}. These leptons, such as atmospheric muons or accelerator muons, if collided by surrounding DM, may be shifted in space, creating a clear signal that can be detected. For simplicity, we illustrate the device in Fig.~\ref{fig:eon} as cubic-like,  with a length of $L \sim 0.1-10$ meters. The device consists of a vacuum region surrounded by tracking detectors. A veto region along the chamber can be defined based on the cross-point of the in and out tracks to suppress further backgrounds.

The maximal recoil energy of a muon is~\cite{Bringmann:2018cvk}
\begin{equation} \label{nonr}
E^{\rm max}_{\rm recoil} = \frac{(2\times {\rm M_D} \times v)^2}{{\rm M_{\mu}}}. 
\end{equation}
The resulting maximum velocity following Eq.~\ref{nonr} is around $10-1000$ km/s, for DM mass $\rm{M_D}\sim$100 (10, 1000) MeV, and the maximum shift distance will be around 1 (0.1, 10) mm, for a cubic device with a length $L\sim 1$ meter. 

Such a device can have a good potential to detect low mass DM. The local density of DM is at the order of $\rho\sim$~0.3 GeV/$\rm{cm}^3$ and with typical velocity of $v=300$\, km/s~\cite{Benito:2019ngh,ParticleDataGroup:2022pth}. The DM flux then can be estimated as $10^7/\rm{M_D}[\rm{GeV}]\rm{cm}^{-2}\rm{s}^{-1}$. When the DM mass $\rm{M_D}\sim$100 (10, 1000) MeV, the flux will be  $10^{8\,(9,\, 7)}\rm{cm}^{-2}\rm{s}^{-1}$.  We can then estimate the DM lepton scattering rate per second as (ignoring the effect of colliding angle below for simplicity): 
\begin{equation}
10^{8\, (9,\,7)}\times \sigma_D \times N_{\mu}, \label{rate}
\end{equation} 
where $N_{\mu}$ represents the muon numbers inside the device and will be discussed more below, and $\sigma_D$ represents the cross section of dark matter scattering with muons.

Notice for high speed muons, it is appropriate to treat DM as frozen in the detector volume ($V$), and the estimated rat per second could be: $\rho V/{\rm M_D} \times \sigma_D \times F_{\mu}$,
where $F_{\mu}$ is for the muon flux ~\cite{Bugaev:1998bi}. 

Furthermore, the dependence on DM velocity distribution has also been explicitly simulated and checked for muon-DM elastic scattering following a truncated Maxwell-Boltzmann distribution~\cite{Wu:2019nhd}, and the effects are found to be small.

\section{Atmospheric muons or Lepton beams} 

Below we discuss about staging possibilities, from smaller to larger size of detectors, using low rate atmospheric muons or high rate accelerator beams.  

Phase I: we can start with a small size device with $L\sim 0.1-1$ meter, and using atmospheric muons to set up a prototype. Based on the muon flux at sea level~\cite{Bugaev:1998bi}, muon numbers inside the device flying from outside at any time, $N_{\mu}$, is around 100 (1000) at the sea (mountain) level. According to Eq.~\ref{rate}, within one year, the sensitivity will be $\sigma_D\sim 10^{-19 (-20,\,-18)}\rm{cm}^{2}$ for $\rm{M_D}=100\, (10,\,1000)$ MeV, estimated simply by assuming $\mathcal{O}(1)$ signal while no backgrounds. 
 
Phase II: we can then interface the device with a high intensity muon beam that could also be used for neutrino oscillation experiments as proposed in~\cite{Ruzi:2023atl}. With a typical bunch density as $10^{12}$/bunch and bunch crossing frequency as $10^5$/sec~\cite{MuonCollider:2022nsa,MuonCollider:2022glg, MuonCollider:2022xlm}, we can get $N_\mu \sim 10^{12}$. According to Eq.~\ref{rate}, within one year, the sensitivity will be $\sigma_D\sim 10^{-27 (-28,\,-26)}\rm{cm}^{2}$ for $\rm{M_D}=100\, (10,\,1000)$ MeV. The results here are comparable with the $M^3$ sensitivity~\cite{Kahn:2018cqs}, $\sigma\sim 10^{-29\sim-30}\rm{cm}^2$, on DM mass range of 10 MeV to 1000 MeV at their phase 1 stage (indeed our proposal can work better for relatively heavy DM). Notice also such a muon beam is collimated with well defined direction, thus the device can be made much smaller in the beam transverse dimensions yet with finer granularity. Similarly, such as device can also be interfaced with a high intensity electron beam, to detect the low mass DM and electron interaction.

\section{Detector Prototype} 

\begin{figure}
    \centering
    \includegraphics[width=1.\columnwidth]{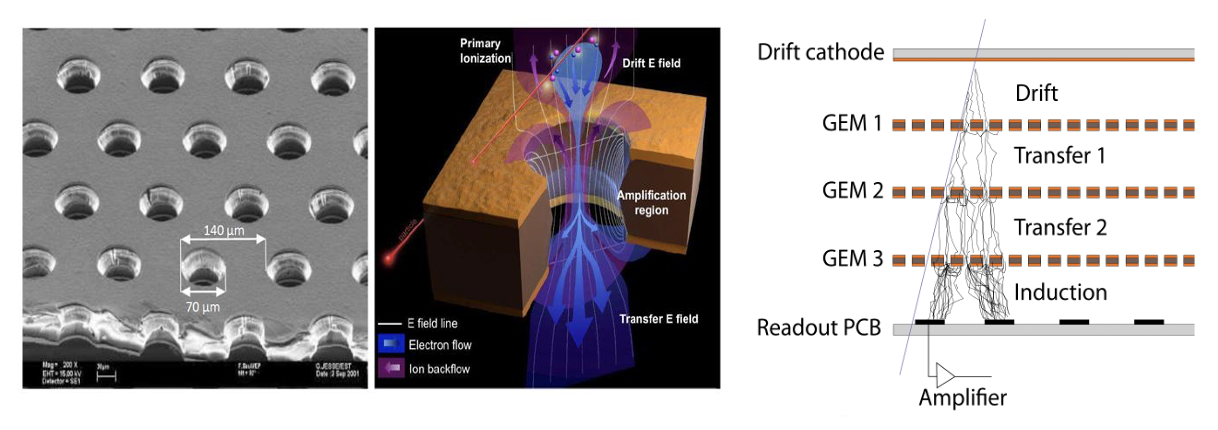}
    \caption{Illustration of the tracking detectors using multi layers Gas Electron Multiplier (GEM)~\cite{gemd} .}
    \label{fig:gem}
\end{figure}

Based on local expertise and the cost-effectiveness of the performance, we plan to exploit Gas Electron Multiplier (GEM) technology for muon tracking, as shown in Fig.~\ref{fig:gem}~\cite{gemd}. GEM is a proven amplification technique for position detection of charged particles in gaseous detectors. GEM has been widely used in various high energy physics experiments, such as the CMS experiment~\cite{Abbas:2022fze,Pellecchia:2022lsd}. We plan to adopt the triple-GEM design from the CMS phase-II detector upgrades: in each chamber, three GEM foils with microscopic holes are stacked between an anode readout board and a cathode drift board. These detectors have shown good performances including detection efficiency, time resolution and spatial resolution. With multiple triple-GEM chambers surrounding sides of a vacuum cube, we aim at a position resolution around 100 microns~\cite{Abbas:2022fze,Pellecchia:2022lsd}. 

The cost of our detector system is mainly driven by the front-end electronics and data-acquisition (DAQ) system.
To provide spatial measurements for muons crossing a vacuum cube with a length of $L\sim 1$ meter, each triple-GEM chamber surrounded the cube sides needs 6400 readout channels, which cost 5,000,000 RMB.
Hence, the total cost of front-end electronics and DAQ system for six triple-GEM chambers is approximately 30,000,000  RMB when working with a $L\sim 1$ meter vacuum cube. The price can be reduced significantly with a smaller cube (notice though the fiducial region and thus sensitivity will get degraded), or a cube with electronics covered only in the top and bottom sides, as a startup prototype.

For a medium or high level vacuum (0.1-100 mPa) with $N_m\sim 10^{19}$ or $10^{15}$ number of molecules per cubic meter~\cite{vacuum}, one can estimate the background annual rate from muons scattering with those molecules. The effects include two parts:  a) muon ionizing atoms by dislodging electrons, and b) muon scattered directly with nuclei. 

\begin{figure}
    \centering
    \includegraphics[width=.4\columnwidth]{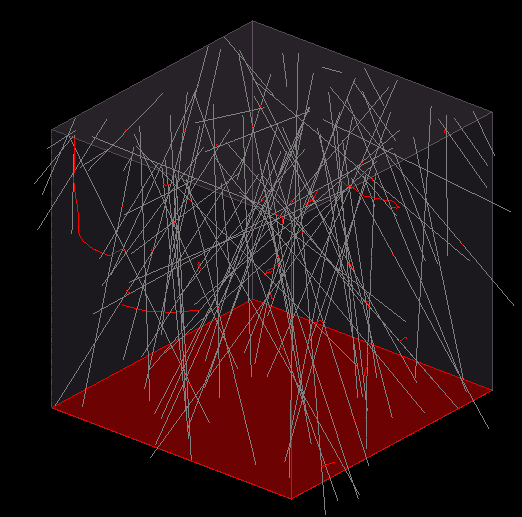}
    \includegraphics[width=.58\columnwidth]{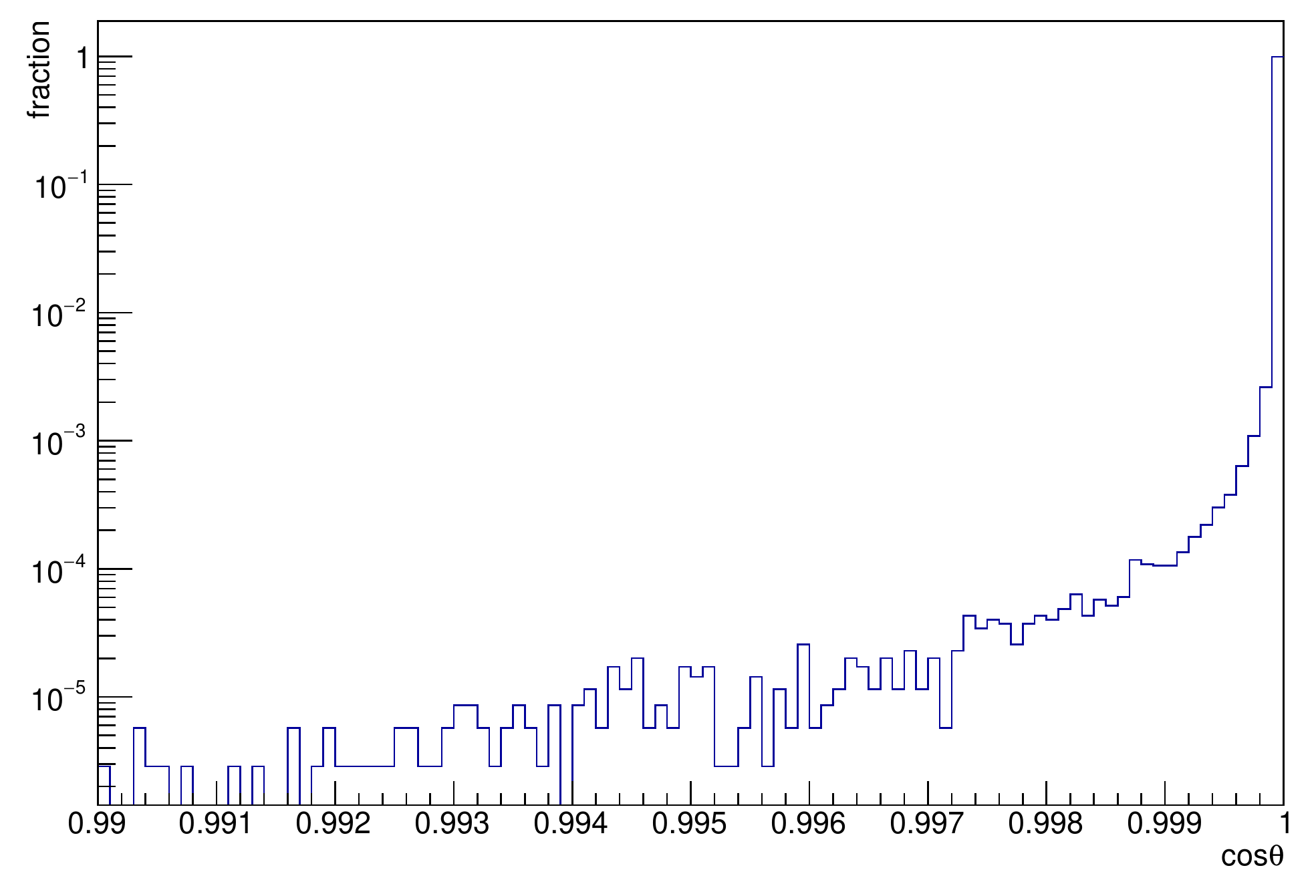}
    \caption{Simulating atmosphere muons passing through a 1 meter cube of air. $|\cos{\theta}|$ distribution is also shown in the right plot, with $\theta$ as the deflected angle between in and out muon tracks. If now working with a high level vacuum which contains around $10^{10}$ times less gas molecules than in the normal air, we estimate that the yearly expected event rates of those muons deflected angles of $|\cos{\theta}|<0.9999999$ to be less than 1-2 per year for the high level vacuum case.}
    \label{fig:g4}
\end{figure}

For a), we simulated with Geant4~\cite{g4} and CRY~\cite{cry} as shown in Fig.~\ref{fig:g4}, where $|\cos{\theta}|$ distribution is also presented, with $\theta$ as the deflected angle between in and out muon tracks.  If now working with a high level vacuum which contains around $10^{10}$ times less gas molecules than in the normal air, we estimate that the yearly expected event rates of those muons deflected angles of $|\cos{\theta}|<0.9999999$ (corresponding maximum kinematic shift is around 0.5 mm which can be well detected with above mentioned triple-GEM device) to be less than 1-2 per year for the high level vacuum case.

For b), we estimate with the following formula:
\begin{equation}
10^{0}\times (\sigma_{\mu p}) \times N_{m} \times 10^7 \sim 10^{-3}, \label{vrate}
\end{equation} 
where $10^{0}~\rm{cm}^{-2}s^{-1}$ is an estimation for the cosmic muon flux, and $\sigma_{\mu p}$ is the muon proton scattering cross section which is roughly at $10^{-30}\rm{cm}^{2}$ level~\cite{mup}.  For a medium or high level vacuum, the event rate is below 0.001 per year.

Correspondingly, the resulted background rate is small and has low impact on DM searches at the Phase I stage of our proposal.  At the Phase II stage, interfacing with high intensity muon beam, the device can be made much smaller in the beam transverse dimensions, thus the number of molecules contained and related contamination can be suppressed to a low level.

Other backgrounds arising at or outside vacuum chamber, can be vetoed effectively by requiring the cross-point of the in and out tracks to be inside the vacuum region, with the veto volume to be optimized through a detailed simulation study. 

Muons bending by the geomagnetic field, ${\rm B_e}\sim 5 \times 10^{-5}$ Tesla, can be estimated with the bending radius:
\begin{equation}
{\rm R}\sim {\rm P}[{\rm GeV}]/\big[0.3\times q[e] \times {\rm B_e[T]}\big] , \label{geom}
\end{equation} 
For GeV scale muons, R is around 1000 km, which is too large to bring significant effects here.

\section{Summary and outlook} 

 We propose a new approach to detect low-mass DM through their scatterings from free leptons. 
 %\sout{Specially, such an experiment can detect DM interacting solely with muons via observing muon deflections caused by scattering off DM particles}. 
 The aim of this proposal is to directly probe muon-philic DM, in a model-independent way. Its complementarity with the muon on target experiment such as $M^3$, is similar with, e.g. XENON/PandaX and ATLAS/CMS on DM searches. Moreover, our proposal can work better for relatively heavy DM such as in the sub-GeV region. 
 
 The proposal can be realized in two phase stages: using energetic and rich-in-flux muons in phase one, and using well collimated muon beams from accelerators in phase two. We start with a small device of a size around 0.1 to 1 meter, using atmospheric muons to set up a prototype. Within only one year of operation in the phase 1 stage, the sensitivity on cross section of DM scattering with muons can already reach $\sigma_D\sim 10^{-19 (-20,\,-18)}\rm{cm}^{2}$ for a dark mater $\rm{M_D}=100\, (10,\,1000)$ MeV. We can then interface the device with a high intensity muon beam of $10^{12}$/bunch. Within one year in phase 2 stage, the sensitivity can reach $\sigma_D\sim 10^{-27 (-28,\,-26)}\rm{cm}^{2}$ for $\rm{M_D}=100\, (10,\,1000)$ MeV. The muons in both cases are interface to a local cost-effective detector based on a Gas Electron Multiplier technology designed for muon tracking. The outgoing muon diverged from its original beam direction is a clear signal of its interaction with low-mass DM particles. Both the air and geomagnetic disturbance on the detection of outgoing muons is negligible because of the vacuumed detector and smallness of the magnetic field of the Earth. Once the results obtained, it can be used to constrain many theoretical models relative to DM sector, especially muon-philic and lepton portal DM.

\appendix
\section{Acknolwledgments}
\begin{acknowledgments}
This work is supported in part by the National Natural Science Foundation of China under Grants No. 12150005, No. 12075004 and No. 12061141002, by MOST under grant No. 2018YFA0403900.
\end{acknowledgments}

\bibliographystyle{ieeetr}
\bibliography{h}

\begin{thebibliography}{10}

\bibitem{Essig:2022dfa}
R.~Essig, G.~K.~Giovanetti, N.~Kurinsky, D.~McKinsey, K.~Ramanathan, K.~Stifter and T.~T.~Yu,
%``Snowmass2021 Cosmic Frontier: The landscape of low-threshold dark matter direct detection in the next decade,''
[arXiv:2203.08297 [hep-ph]].
%\cite{Bai:2014osa}
\bibitem{Harris:2022vnx}
P.~Harris, P.~Schuster and J.~Zupan,
%``Snowmass White Paper: New flavors and rich structures in dark sectors,''
[arXiv:2207.08990 [hep-ph]].
\bibitem{Bai:2014osa}
Y.~Bai and J.~Berger,
%``Lepton Portal Dark Matter,''
JHEP \textbf{08}, 153 (2014)
doi:10.1007/JHEP08(2014)153
[arXiv:1402.6696 [hep-ph]].
%106 citations counted in INSPIRE as of 11 Mar 2023

\bibitem{migdal} M. Ibe, W. Nakano, Y. Shoji, and K. Suzuki, JHEP 03,
194 (2018), arXiv:1707.07258 [hep-ph]
\bibitem{EDELWEISS}
Q. Arnaud \textit{et al.} (EDELWEISS), Phys. Rev. Lett. 125,
141301 (2020), arXiv:2003.01046 [astro-ph.GA]
\bibitem{SENSEI}
L. Barak \textit{et al.} (SENSEI), Phys. Rev. Lett. 125, 171802
(2020), arXiv:2004.11378 [astro-ph.CO].
\bibitem{Liao:2021npo}
J.~Liao, Y.~Gao, Z.~Liang, Z.~Peng, Z.~Ouyang, L.~Zhang, L.~Zhang and J.~Zhou,
%``A low-mass dark matter project, ALETHEIA: A Liquid hElium Time projection cHambEr In dArk matter,''
[arXiv:2103.02161 [astro-ph.IM]].
\bibitem{Liao:2022zqg}
J.~Liao, Y.~Gao, Z.~Jiang, Z.~Liang, Z.~OuYang, Z.~Peng, F.~Zhang, L.~Zhang and J.~Zhou,
%``ALETHEIA: Hunting for Low-mass Dark Matter with Liquid Helium TPCs,''
doi:10.1140/epjp/s13360-023-03747-2
[arXiv:2209.02320 [astro-ph.IM]].
\bibitem{Super-Kamiokande:2022ncz}
K.~Abe \textit{et al.} [Super-Kamiokande],
%``Search for Cosmic-Ray Boosted Sub-GeV Dark Matter Using Recoil Protons at Super-Kamiokande,''
Phys. Rev. Lett. \textbf{130}, no.3, 031802 (2023)
doi:10.1103/PhysRevLett.130.031802
[arXiv:2209.14968 [hep-ex]].
\bibitem{PandaX-II:2021kai}
X.~Cui \textit{et al.} [PandaX-II],
%``Search for Cosmic-Ray Boosted Sub-GeV Dark Matter at the PandaX-II Experiment,''
Phys. Rev. Lett. \textbf{128}, no.17, 171801 (2022)
doi:10.1103/PhysRevLett.128.171801
[arXiv:2112.08957 [hep-ex]].
\bibitem{Bringmann:2018cvk}
T.~Bringmann and M.~Pospelov,
%``Novel direct detection constraints on light dark matter,''
Phys. Rev. Lett. \textbf{122}, no.17, 171801 (2019)
doi:10.1103/PhysRevLett.122.171801
[arXiv:1810.10543 [hep-ph]].
\bibitem{Plestid:2020kdm}
R.~Plestid, V.~Takhistov, Y.~D.~Tsai, T.~Bringmann, A.~Kusenko and M.~Pospelov,
%``New Constraints on Millicharged Particles from Cosmic-ray Production,''
Phys. Rev. D \textbf{102}, 115032 (2020)
doi:10.1103/PhysRevD.102.115032
[arXiv:2002.11732 [hep-ph]].
\bibitem{Hu:2016xas}
P.~K.~Hu, A.~Kusenko and V.~Takhistov,
%``Dark Cosmic Rays,''
Phys. Lett. B \textbf{768}, 18-22 (2017)
doi:10.1016/j.physletb.2017.02.035
[arXiv:1611.04599 [hep-ph]].

\bibitem{Elor:2021swj}
G.~Elor, R.~McGehee and A.~Pierce,
%``Maximizing Direct Detection with Highly Interactive Particle Relic Dark Matter,''
Phys. Rev. Lett. \textbf{130}, no.3, 031803 (2023)
doi:10.1103/PhysRevLett.130.031803
[arXiv:2112.03920 [hep-ph]].

%\cite{Essig:2012yx}
\bibitem{Essig:2012yx}
R.~Essig, A.~Manalaysay, J.~Mardon, P.~Sorensen and T.~Volansky,
%``First Direct Detection Limits on sub-GeV Dark Matter from XENON10,''
Phys. Rev. Lett. \textbf{109}, 021301 (2012)
doi:10.1103/PhysRevLett.109.021301
[arXiv:1206.2644 [astro-ph.CO]].
%413 citations counted in INSPIRE as of 11 Mar 2023

%\cite{Battaglieri:2020lds}
\bibitem{Battaglieri:2020lds}
M.~Battaglieri, P.~Bisio, M.~Bond\'\i{}, A.~Celentano, P.~L.~Cole, M.~De Napoli, R.~De Vita, L.~Marsicano, G.~Ottonello and F.~Parodi, \textit{et al.}
%``The BDX-MINI detector for Light Dark Matter search at JLab,''
Eur. Phys. J. C \textbf{81}, no.2, 164 (2021)
doi:10.1140/epjc/s10052-021-08957-5
[arXiv:2011.10532 [physics.ins-det]].
%6 citations counted in INSPIRE as of 11 Mar 2023

%\cite{Berlin:2020uwy}
\bibitem{Berlin:2020uwy}
A.~Berlin, P.~deNiverville, A.~Ritz, P.~Schuster and N.~Toro,
%``Sub-GeV dark matter production at fixed-target experiments,''
Phys. Rev. D \textbf{102}, no.9, 095011 (2020)
doi:10.1103/PhysRevD.102.095011
[arXiv:2003.03379 [hep-ph]].
%22 citations counted in INSPIRE as of 11 Mar 2023

%\cite{LDMX:2018cma}
\bibitem{LDMX:2018cma}
T.~\r{A}kesson \textit{et al.} [LDMX],
%``Light Dark Matter eXperiment (LDMX),''
[arXiv:1808.05219 [hep-ex]].
%155 citations counted in INSPIRE as of 11 Mar 2023

%\cite{DarkSide:2022knj}
\bibitem{DarkSide:2022knj}
P.~Agnes \textit{et al.} [DarkSide],
%``Search for Dark Matter Particle Interactions with Electron Final States with DarkSide-50,''
Phys. Rev. Lett. \textbf{130}, no.10, 101002 (2023)
doi:10.1103/PhysRevLett.130.101002
[arXiv:2207.11968 [hep-ex]].
%12 citations counted in INSPIRE as of 11 Mar 2023

%\cite{AlAli:2021let}
\bibitem{AlAli:2021let}
H.~Al Ali, N.~Arkani-Hamed, I.~Banta, S.~Benevedes, D.~Buttazzo, T.~Cai, J.~Cheng, T.~Cohen, N.~Craig and M.~Ekhterachian, \textit{et al.}
%``The muon Smasher\textquoteright{}s guide,''
Rept. Prog. Phys. \textbf{85}, no.8, 084201 (2022)
doi:10.1088/1361-6633/ac6678
[arXiv:2103.14043 [hep-ph]].
%82 citations counted in INSPIRE as of 11 Mar 2023

%\cite{Muong-2:2021ojo}
\bibitem{Muong-2:2021ojo}
B.~Abi \textit{et al.} [Muon g-2],
%``Measurement of the Positive Muon Anomalous Magnetic Moment to 0.46 ppm,''
Phys. Rev. Lett. \textbf{126}, no.14, 141801 (2021)
doi:10.1103/PhysRevLett.126.141801
[arXiv:2104.03281 [hep-ex]].
%1232 citations counted in INSPIRE as of 11 Mar 2023

%\cite{Forbes:2022bvo}
\bibitem{Forbes:2022bvo}
D.~Forbes, C.~Herwig, Y.~Kahn, G.~Krnjaic, C.~Mantilla Suarez, N.~Tran and A.~Whitbeck,
%``New Searches for Muonphilic Particles at Proton Beam Dump Spectrometers,''
[arXiv:2212.00033 [hep-ph]].
%0 citations counted in INSPIRE as of 12 Mar 2023.

%\cite{Gninenko:2014pea}
\bibitem{Gninenko:2014pea}
S.~N.~Gninenko, N.~V.~Krasnikov and V.~A.~Matveev,
%``Muon g-2 and searches for a new leptophobic sub-GeV dark boson in a missing-energy experiment at CERN,''
Phys. Rev. D \textbf{91}, 095015 (2015)
doi:10.1103/PhysRevD.91.095015
[arXiv:1412.1400 [hep-ph]].
%103 citations counted in INSPIRE as of 12 Mar 2023

%\cite{Chen:2018vkr}
\bibitem{Chen:2018vkr}
C.~Y.~Chen, J.~Kozaczuk and Y.~M.~Zhong,
%``Exploring leptophilic dark matter with NA64-$\mu$,''
JHEP \textbf{10}, 154 (2018)
doi:10.1007/JHEP10(2018)154
[arXiv:1807.03790 [hep-ph]].
%34 citations counted in INSPIRE as of 12 Mar 2023

%\cite{Kahn:2018cqs}
\bibitem{Kahn:2018cqs}
Y.~Kahn, G.~Krnjaic, N.~Tran and A.~Whitbeck,
JHEP \textbf{09}, 153 (2018)
doi:10.1007/JHEP09(2018)153
[arXiv:1804.03144 [hep-ph]].
%85 citations counted in INSPIRE as of 12 Mar 2023


%\cite{Chen:2017awl}
\bibitem{Chen:2017awl}
C.~Y.~Chen, M.~Pospelov and Y.~M.~Zhong,
%``Muon Beam Experiments to Probe the Dark Sector,''
Phys. Rev. D \textbf{95}, no.11, 115005 (2017)
doi:10.1103/PhysRevD.95.115005
[arXiv:1701.07437 [hep-ph]].
%62 citations counted in INSPIRE as of 12 Mar 2023

%\cite{Cesarotti:2022ttv}
\bibitem{Cesarotti:2022ttv}
C.~Cesarotti, S.~Homiller, R.~K.~Mishra and M.~Reece,
%``Probing New Gauge Forces with a High-Energy Muon Beam Dump,''
Phys. Rev. Lett. \textbf{130}, no.7, 071803 (2023)
doi:10.1103/PhysRevLett.130.071803
[arXiv:2202.12302 [hep-ph]].
%16 citations counted in INSPIRE as of 12 Mar 2023

\bibitem{Buchmueller:2017qhf}
O.~Buchmueller, C.~Doglioni and L.~T.~Wang,
%``Search for dark matter at colliders,''
Nature Phys. \textbf{13}, no.3, 217-223 (2017)
doi:10.1038/nphys4054
[arXiv:1912.12739 [hep-ex]].


\bibitem{Benito:2019ngh}
M.~Benito, A.~Cuoco and F.~Iocco,
%``Handling the Uncertainties in the Galactic Dark Matter Distribution for Particle Dark Matter Searches,''
JCAP \textbf{03}, 033 (2019)
doi:10.1088/1475-7516/2019/03/033
[arXiv:1901.02460 [astro-ph.GA]].

\bibitem{ParticleDataGroup:2022pth}
R.~L.~Workman \textit{et al.} [Particle Data Group],
%``Review of Particle Physics,''
PTEP \textbf{2022}, 083C01 (2022)
doi:10.1093/ptep/ptac097

\bibitem{Bugaev:1998bi}
E.~V.~Bugaev, A.~Misaki, V.~A.~Naumov, T.~S.~Sinegovskaya, S.~I.~Sinegovsky and N.~Takahashi,
%``Atmospheric muon flux at sea level, underground and underwater,''
Phys. Rev. D \textbf{58}, 054001 (1998)
doi:10.1103/PhysRevD.58.054001
[arXiv:hep-ph/9803488 [hep-ph]].

\bibitem{Wu:2019nhd}
Y.~Wu, K.~Freese, C.~Kelso, P.~Stengel and M.~Valluri,
%``Uncertainties in Direct Dark Matter Detection in Light of Gaia's Escape Velocity Measurements,''
JCAP \textbf{10}, 034 (2019)
doi:10.1088/1475-7516/2019/10/034
[arXiv:1904.04781 [hep-ph]].

%\cite{Ruzi:2023atl}
\bibitem{Ruzi:2023atl}
A.~Ruzi, T.~Yang, D.~Fu, S.~Qian, L.~Gao and Q.~Li,
%``Muon Beam for Neutrino CP Violation: connecting energy and neutrino frontiers,''
[arXiv:2301.02493 [hep-ph]].
%0 citations counted in INSPIRE as of 14 Mar 2023

\bibitem{MuonCollider:2022nsa}
D.~Stratakis \textit{et al.} [Muon Collider],
%``A Muon Collider Facility for Physics Discovery,''
[arXiv:2203.08033 [physics.acc-ph]].

\bibitem{MuonCollider:2022glg}
S.~Jindariani \textit{et al.} [Muon Collider],
%``Promising Technologies and R\&D Directions for the Future Muon Collider Detectors,''
[arXiv:2203.07224 [physics.ins-det]].

\bibitem{MuonCollider:2022xlm}
J.~de Blas \textit{et al.} [Muon Collider],
%``The physics case of a 3 TeV muon collider stage,''
[arXiv:2203.07261 [hep-ph]].

\bibitem{gemd}
https://ep-news.web.cern.ch/content/gems-cms

\bibitem{Abbas:2022fze}
M.~Abbas, M.~Abbrescia, H.~Abdalla, A.~Abdelalim, S.~AbuZeid, A.~Agapitos, A.~Ahmad, A.~Ahmed, W.~Ahmed and C.~Aim\`e, \textit{et al.}
%``Quality control of mass-produced GEM detectors for the CMS GE1/1 muon upgrade,''
Nucl. Instrum. Meth. A \textbf{1034}, 166716 (2022)
doi:10.1016/j.nima.2022.166716
[arXiv:2203.12037 [physics.ins-det]].

\bibitem{Pellecchia:2022lsd}
A.~Pellecchia \textit{et al.} [CMS Muon],
%``Performance of triple-GEM detectors for the CMS Phase-2 upgrade measured in test beam,''
Nucl. Instrum. Meth. A \textbf{1046}, 167618 (2023)
doi:10.1016/j.nima.2022.167618
[arXiv:2207.09906 [physics.ins-det]].

\bibitem{vacuum} https://www.mks.com/n/vacuum-basics
\bibitem{g4} S. Agostinelli et al. [GEANT4], Nucl. Instrum. Meth. A
506, 250-303 (2003) doi:10.1016/S0168-9002(03)01368-8
\bibitem{cry} Hagmann, C., Lange, D., Verbeke, J., Wright, D., 2012. Cosmic-ray Shower Library (CRY),
Lawrence Livermore NationalLaboratory, https://nuclear.llnl.gov/simulation/
\bibitem{mup} L. Camilleri, \textit{et al.} Phys. Rev. Lett. 23, 153 (1969).


\end{thebibliography}
\end{document}